\documentclass[superscriptaddress,reprint,aps,prl]{revtex4-1}

\usepackage{amssymb, amsmath}
\usepackage{graphicx}
\usepackage{dcolumn}
\usepackage{bm}

\begin{document}


\title{Magnetostriction and magneto-structural domains in antiferromagnetic \mbox{YBa$_{2}$Cu$_{3}$O$_{6}$}}
\author{B.~N\'afr\'adi}
\affiliation{Max-Planck-Institut f\"ur Festk\"orperforschung, Heisenbergstra\ss e 1, D-70569 Stuttgart, Germany}
\author{T.~Keller}
\affiliation{Max-Planck-Institut f\"ur Festk\"orperforschung, Heisenbergstra\ss e 1, D-70569 Stuttgart, Germany}
\affiliation{Max Planck Society Outstation at the Heinz Maier-Leibnitz Zentrum (MLZ), D-85748 Garching, Germany}
\author{F.~Hardy}
\affiliation{Institut f\"ur Festk\"orperphysik, Karlsruher Institut f\"ur Technlogie (KIT), D-76344 Eggenstein-Leopoldshafen, Germany}
\author{C.~Meingast}
\affiliation{Institut f\"ur Festk\"orperphysik, Karlsruher Institut f\"ur Technlogie (KIT), D-76344 Eggenstein-Leopoldshafen, Germany}
\author{A.~Erb}
\affiliation{Walter Meissner Institut f\"ur Tieftemperaturforschung D-85748 Garching, Germany}
\author{B.~Keimer}
\affiliation{Max-Planck-Institut f\"ur Festk\"orperforschung, Heisenbergstra\ss e 1, D-70569 Stuttgart, Germany}

\date{\today}

\begin{abstract}
We have used high-resolution neutron Larmor diffraction and capacitative dilatometry to investigate spontaneous and forced magnetostriction in undoped, antiferromagnetic YBa$_2$Cu$_3$O$_{6.0}$, the parent compound of a prominent family of high-temperature superconductors.
Upon cooling below the N\'eel temperature, $T_N = 420$~K, Larmor diffraction reveals the formation of magneto-structural domains of characteristic size $\sim 240$~nm.
In the antiferromagnetic state, dilatometry reveals a minute ($4 \times 10^{-6}$) orthorhombic distortion of the crystal lattice in external magnetic fields.
We attribute these observations to exchange striction and spin-orbit coupling induced magnetostriction, respectively, and show that they have an important influence on the thermal and charge transport properties of undoped and lightly doped cuprates.
\end{abstract}


\maketitle

Correlated-electron systems exhibit multiple collective ordering phenomena whose interdependence and competition are subjects of intense current research.
The macroscopic properties of materials with strongly correlated electrons are influenced not only by atomic-scale correlations, but also by emergent domain structures on nanoscopic and mesoscopic length scales \cite{Dagotto}.
Recent advances in research on some of the most prominent correlated-electron materials, the cuprate high-temperature superconductors, \cite{Keimer} have reinforced efforts to establish quantitative links between the doping dependent spin and charge correlations and the thermodynamic and transport properties \cite{Capati,Fernandes,Senthil}.
These efforts are, however, complicated by the presence of defects and associated strains of the crystal lattice, which are invariably associated with doping and strongly affect the mesoscopic organization of the electron system \cite{Zeljkovic}.
Recent examples include magnetic hysteresis phenomena \cite{Kapitulnik,Shi} and charge density wave pinning \cite{Achkar,LeTacon,Julien} in moderately doped superconducting cuprates, whose origins have not yet been conclusively identified.

To provide a solid basis for the investigation of doped high-temperature superconductors, it is important to establish a firm understanding of electronic correlations and their coupling to the crystal lattice in the undoped, largely defect-free parent compounds that exhibit antiferromagnetic long-range order.
Although the atomic-scale spin correlations of undoped cuprates are well understood, there is little direct information on antiferromagnetic domain structures and associated lattice strains, despite indications that they profoundly affect the charge \cite{Ando,Cimpoiasu} and heat \cite{Hess,Hofmann} transport properties and may act as seeds for mesoscopic inhomogeneities in doped compounds \cite{Keimer,Zeljkovic}.
In particular, an anomalous magnetoresistance has been reported for lightly doped, antiferromagnetic YBa$_2$Cu$_3$O$_{6+\delta}$, \cite{Ando,Cimpoiasu} a material that has served as a model compound for recent research on high-temperature superconductivity \cite{Keimer}.
The magnetoresistance in the CuO$_2$-planes was found to exhibit a ``$d$-wave'' symmetry upon rotation of the magnetic field in this plane, that is, the resistance increases (decreases) when the magnetic field is parallel (perpendicular) to the current flow \cite{Ando,Cimpoiasu}.
This finding was unexpected, because at low doping levels the crystal lattice is believed to be tetragonal \cite{Jorgensen}.
In this lattice structure, the two orthogonal $a$ axes in the CuO$_2$ planes are equivalent, and current flow along both axes should be identical.
Ando \emph{et al.} \cite{Ando} attributed the anomalous magnetoresistance to the magnetic-field-induced reorientation of charge stripes that locally break the tetragonal symmetry of the CuO$_2$ planes.
Related ideas have also been discussed for other families of cuprate superconductors \cite{Keimer}.
An alternative model \cite{Cimpoiasu,Janossy1,Janossy2,Gomonaj,Janossy3} invokes antiferromagnetic domains that are accompanied by a small orthorhombic lattice distortion due to magnetostriction and are reoriented by the magnetic field.
The orthorhombic distortion was estimated \cite{Cimpoiasu} as $a/b-1 \sim 6 \times 10^{-6}$, a value too small to be directly observed by x-ray or neutron diffraction techniques.
Likewise, direct evidence of the purported charge-stripe or magneto-elastic domains has thus far not been reported for undoped and lightly doped YBa$_2$Cu$_3$O$_{6+\delta}$.

In the present work, we used high-resolution neutron Larmor diffraction to directly measure the magneto-structural domain size, and capacitative dilatometry to determine the minute orthorhombicity in the antiferromagnetic state by field-aligning the magnetic domains.
We discuss these phenomena in terms of different mechanisms of magnetostriction, and compare the results quantitatively with heat and charge transport data on undoped and lightly doped cuprates.
The methodology we introduce provides interesting perspectives for the investigation of domain structures associated with charge density waves in more highly doped cuprates, and with electronic ordering phenomena in other correlated-electron materials such as the iron pnictides and chalcogenides.

\begin{figure}
	\centering
	\includegraphics[width=\columnwidth]{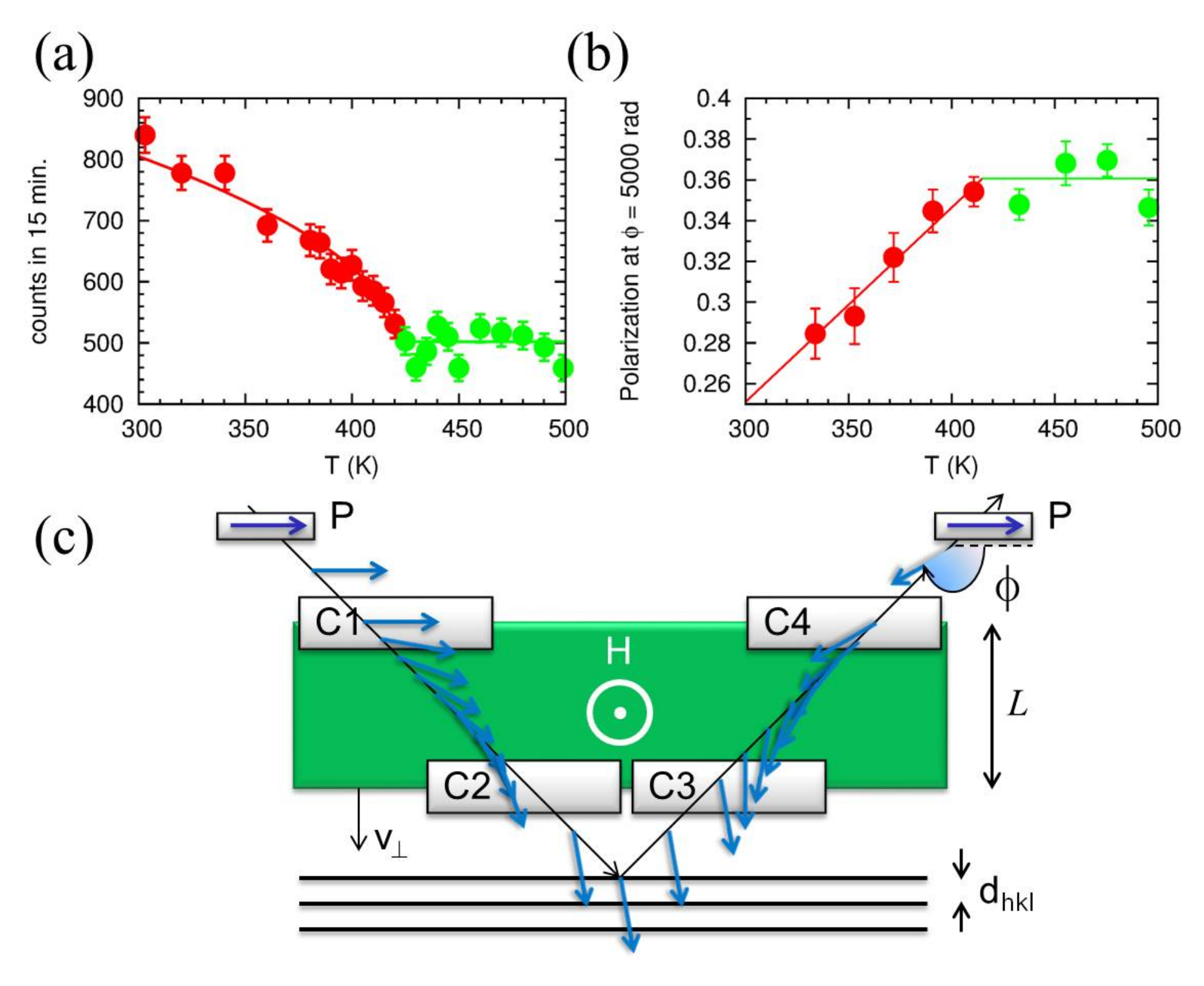}
    \caption{(color online)
    (a) Temperature dependence of the $(\frac{1}{2}\:\frac{1}{2}\:5)$ antiferromagnetic Bragg peak intensity measured by neutron diffraction from YBa$_2$Cu$_3$O$_{6.0}$. The line is a guide to the eye. Red and green symbols indicate temperatures below and above $T_N = 420$~K, respectively. (b)
    Neutron beam polarization $P$ measured at the $(2\:0\:0)$ Bragg reflection for Larmor phase $\phi=5000$ rad (see Fig. 2). The reduction of $P$ below $T_N$ indicates a reduction of the structural domain size. (c) Sketch of the Larmor diffraction method. The radio frequency coils (C1-C4) act on the neutron spins (blue) in the same way as an effective static magnetic field $H$ (green). \label{fig:1}}		
\end{figure}

The experiments were carried out on high-quality YBa$_2$Cu$_3$O$_{6.0}$ single crystals of typical size $1 \times 1 \times 0.1$~$\text{mm}^3$ and mosaicity $\leq 0.1^\circ$, which were grown by a flux method \cite{Erb}.
For the dilatometry measurements, a single specimen was mounted in a capacitance dilatometer \cite{Meingast}, such that the expansion of the $a$-axis was measured. A small force of $20~\text{N}$ along the $a$-axis was applied to hold the crystal, resulting in a uniaxial pressure of $\simeq 200~\text{MPa}$.
The dilatometer was installed in three different orientations in a $10~\text{T}$ magnet to apply the field along the crystallographic $a$, $b$, or $c$-axes.
For the neutron scattering experiments, fifteen crystals of total mass 0.1~g were co-aligned with combined mosaicity $\sim 1^\circ$.
The temperature dependence of the magnetic $(0.5\:0.5\:5)$ Bragg peak intensity (Fig.~\ref{fig:1}a) shows a N\'eel temperature of $T_N=420~\text{K}$, corresponding to full oxygen stoichiometry ($6.0$ oxygen atoms per formula unit) \cite{Shamoto}.

The neutron Larmor diffraction experiments were conducted at the TRISP spectrometer at the Heinz-Maier-Leibnitz Zentrum in Garching
\cite{Keller}.
The basic principle of a Larmor diffractometry (LD) is shown in Fig.~\ref{fig:1}c. \cite{Rekveldt}.
A spin-polarized neutron beam crosses a uniform magnetic field $H$ twice, before and after being diffracted at lattice planes with spacing $d_{\text{hkl}}=2\pi/G_{\text{hkl}}$, where $G_{\text{hkl}}$ is the reciprocal lattice vector.
The boundaries of $H$ are aligned parallel to the lattice planes.
Inside the field, the neutron spins precess with the Larmor frequency $\omega_L=2\pi\gamma H$, where $\gamma$
is the neutron's gyromagnetic ratio.
The total precession angle is $\phi = \omega_L t$, where $t=2L/v_{\perp}$ is the time the neutron spends in the field.
$t$ only depends on the velocity component $v_{\perp}=(\hbar/m) G_{\text{hkl}}/2$, which is independent of the Bragg angle ($m$ is the neutron mass).
The total phase $\phi=2m/(\pi\hbar)\omega_L d_{\text{hkl}}$ is thus a measure for $d_{\text{hkl}}$.
A broadening of the Bragg reflection $\Delta G_{\text{hkl}}$  gives rise to a linear variation of the Larmor phase $\Delta \phi / \phi = \epsilon_\text{hkl}$, with $\epsilon_\text{hkl}=\Delta G_{\text{hkl}} / G_{\text{hkl}}$.
The beam polarization $P(\phi)$ is then the Fourier transform of the momentum-space profile $f(\epsilon_\text{hkl})$ of the Bragg reflection, so that the width of $P$ is the inverse of the width of $f$:

\begin{equation} \label{P_phi}
P(\phi) = \int f(\epsilon_\text{hkl}) \cos(\phi \cdot\epsilon_\text{hkl})\: d\epsilon_\text{hkl}
\end{equation}

Conventional diffractometers are based on measurements of the Bragg angle, where the resolution is limited by the collimation and the monochromaticity of the neutron beam.
The resolution of LD, on the other hand, is limited by the relative error $\delta \phi/\phi$.
The leading contribution to $\delta \phi$ are fluctuations of $H$, which can be strongly reduced by replacing the static field by four radio-frequency spin-flip coils C1-C4 (Fig.~\ref{fig:1}c).
In this way, the momentum-space resolution can be enhanced by about two orders of magnitude \cite{Golub}.

\begin{figure}
	\includegraphics[width=\columnwidth]{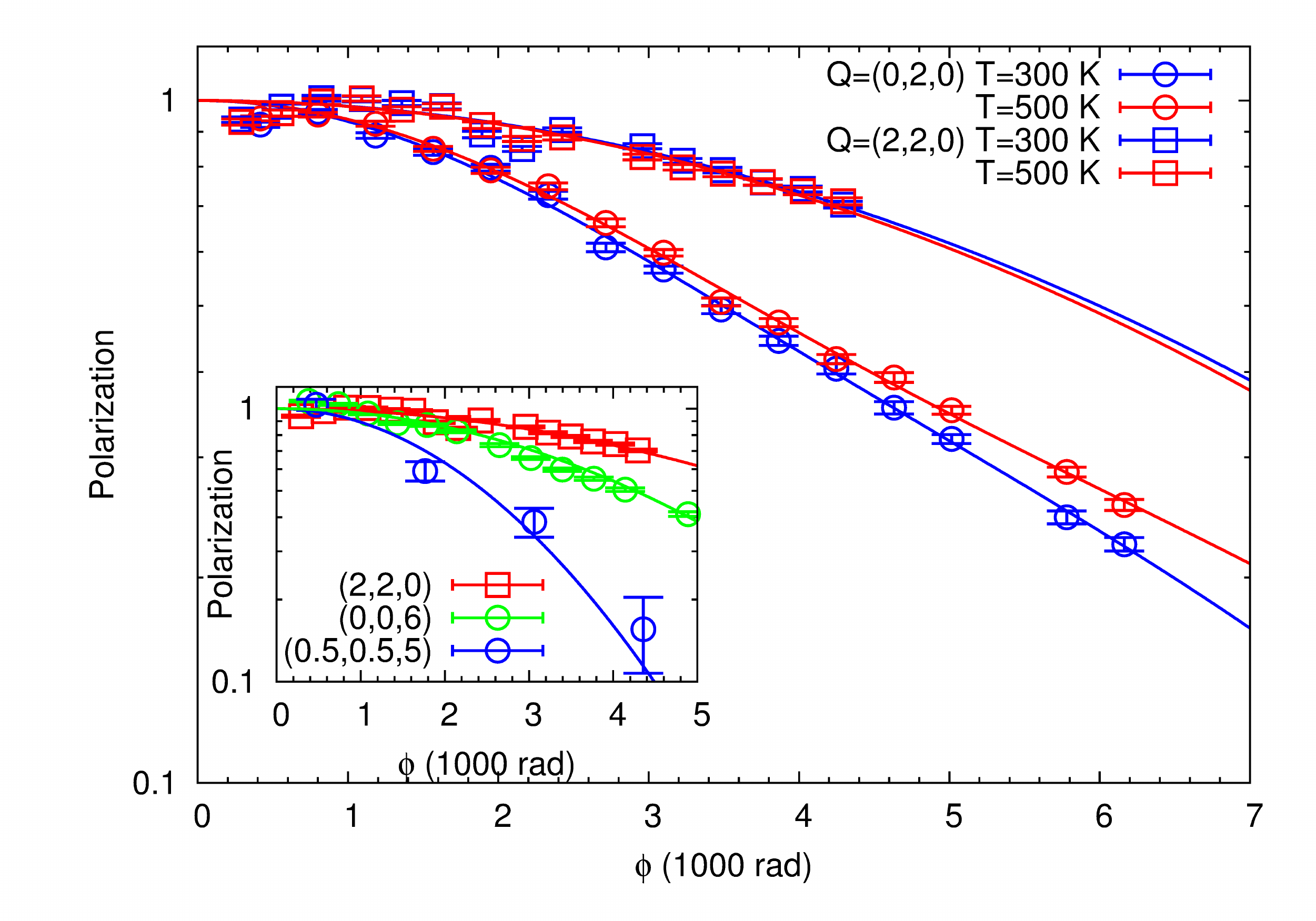}
    \caption{(color online) Neutron beam polarization $P$ versus Larmor phase $\phi$ at $T = 300$ and 500~K for the $(2\:2\:0)$ and $(2\:0\:0)$ nuclear Bragg peaks. $P(\phi=0)$ is normalized to 1. Lines are the results of fits to Gaussian peak profiles (see text). Inset: $P(\phi)$ at $T=40 $~K for the $(\frac{1}{2}\:\frac{1}{2}\:5)$ antiferromagnetic Bragg reflection (blue), compared to the $(2\:2\:0)$ and $(0\:0\:6)$ nuclear Bragg reflections (red and green, respectively).\label{fig:2}}
\end{figure}

Figure~\ref{fig:2} shows $P(\phi)$ profiles for several nuclear and magnetic Bragg reflections.
The instrumental resolution was taken into account by normalizing the profiles to the one obtained from a perfect germanium crystal.
For clarity, the data are displayed after normalization to $P(0) =1$.
To extract the widths of Bragg reflections from the LD data, the $P(\phi)$ curves were fitted to Eq.~\ref{P_phi} with Gaussian peak profiles $f(\epsilon_\text{hkl})$ (lines in Fig.~\ref{fig:2}).
The widths of the $(2\:0\:0)$ and $(2\:2\:0)$ nuclear Bragg peaks determined in this way are quite different (Fig.~\ref{fig:2}).
For $T > T_N$, the width of the $(2\:0\:0)$ reflection, $\epsilon = 5.2 \times 10^{-4}$, translates into a characteristic length $L_{\parallel} = 370$~nm, and the ratio of 1.4 between the widths of the $(2\:2\:0)$ and $(2\:0\:0)$ reflections matches the ratio of their respective reciprocal lattice vectors.
The LD data are thus consistent with square-shaped structural mosaic blocks of characteristic size $L_{\parallel}$ along the CuO$_2$ planes.
The domain size along the $c$-axis extracted from the  $(0\:0\:6)$ reflection (inset in Fig.~\ref{fig:2}) is $L_{\perp} \sim 390$~nm.
Possible origins of structural domain formation include a small number of residual impurities (such as interstitial oxygen) and associated microstrains.
A detailed analysis of the lattice defects in the paramagnetic state will require a survey of multiple Bragg reflections and is beyond the scope of this paper, which is focused on the influence of the electronically driven antiferromagnetic transition on the lattice structure.

To this end, we have carefully monitored the evolution of the $P(\phi)$ profiles across the antiferromagnetic phase transition (Fig.~\ref{fig:2}). The width of the $(2\:0\:0)$ reflection for $T < T_N$ translates into a characteristic domain size of $L_{\parallel} \sim 340$ nm, about 10\% smaller than in the paramagnetic state. The $T$-dependence of the profiles (Fig.~\ref{fig:1}b) demonstrates that the broadening of $P(\phi)$ and the reduction of $L_{\parallel}$ set in at $T = T_N$. Within the experimental error, the ratio of the $(2\:0\:0)$ and $(2\:2\:0)$ widths is preserved upon cooling across $T_N$ (Fig.~\ref{fig:2}), indicating shape-preserving shrinkage of the structural mosaic blocks as the spin fluctuations are arrested in the antiferromagnetic state.

The anomalous broadening of the $P(\phi)$ profiles is a manifestation of coupling between the antiferromagnetic order parameter and the crystal lattice. In rare-earth antiferromagnets, magneto-structural interactions have been detected through anomalies in the thermal expansion at the N\'eel temperature, and were attributed to the dependence of the exchange interactions on the distance between the magnetic ions (``exchange striction'') \cite{Doerr}. In the cuprates, however, such anomalies are much harder to recognize because of the quasi-two-dimensional nature of the magnetism, which implies that the spin correlations in the CuO$_2$ planes are already well developed for $T = T_N$. \cite{Nagel} Our data establish Larmor diffraction as an alternative, highly sensitive probe of magnetostriction in this situation. Following prior theoretical work \cite{Doerr}, the reduction of the structural domain size at $T_N$ observed in YBa$_2$Cu$_3$O$_6$ can be understood as a consequence of exchange striction, which stiffens the crystal lattice so that it can less easily accommodate strains from residual impurities and defects. The fact that the shape of the mosaic blocks remains unchanged at the N\'eel transition agrees with the observation that the exchange Hamiltonian has the same (tetragonal) symmetry as the crystal lattice (apart from the minute effect of the spin-orbit interaction, to be discussed below). In the iron arsenides, by contrast, the symmetry of the magnetic bond network differs from the one of the crystal lattice in the paramagnetic state, giving rise to a sequence of distinct structural and magnetic phase transitions.

The width of the LD profile of the antiferromagnetic Bragg reflection $(\frac{1}{2}\:\frac{1}{2}\:5)$ is comparable to, but somewhat larger than those of the structural reflections (Fig.~\ref{fig:2}), consistent with the expectation that structural domain boundaries resulting from magnetostriction will usually disrupt magnetic order \cite{note}. The spatially averaged antiferromagnetic domain size of 240~nm is quite comparable to the magnetic domain size measured by LD in classical antiferromagnets \cite{Bayrakci}.

\begin{figure}
	\includegraphics[width=\columnwidth]{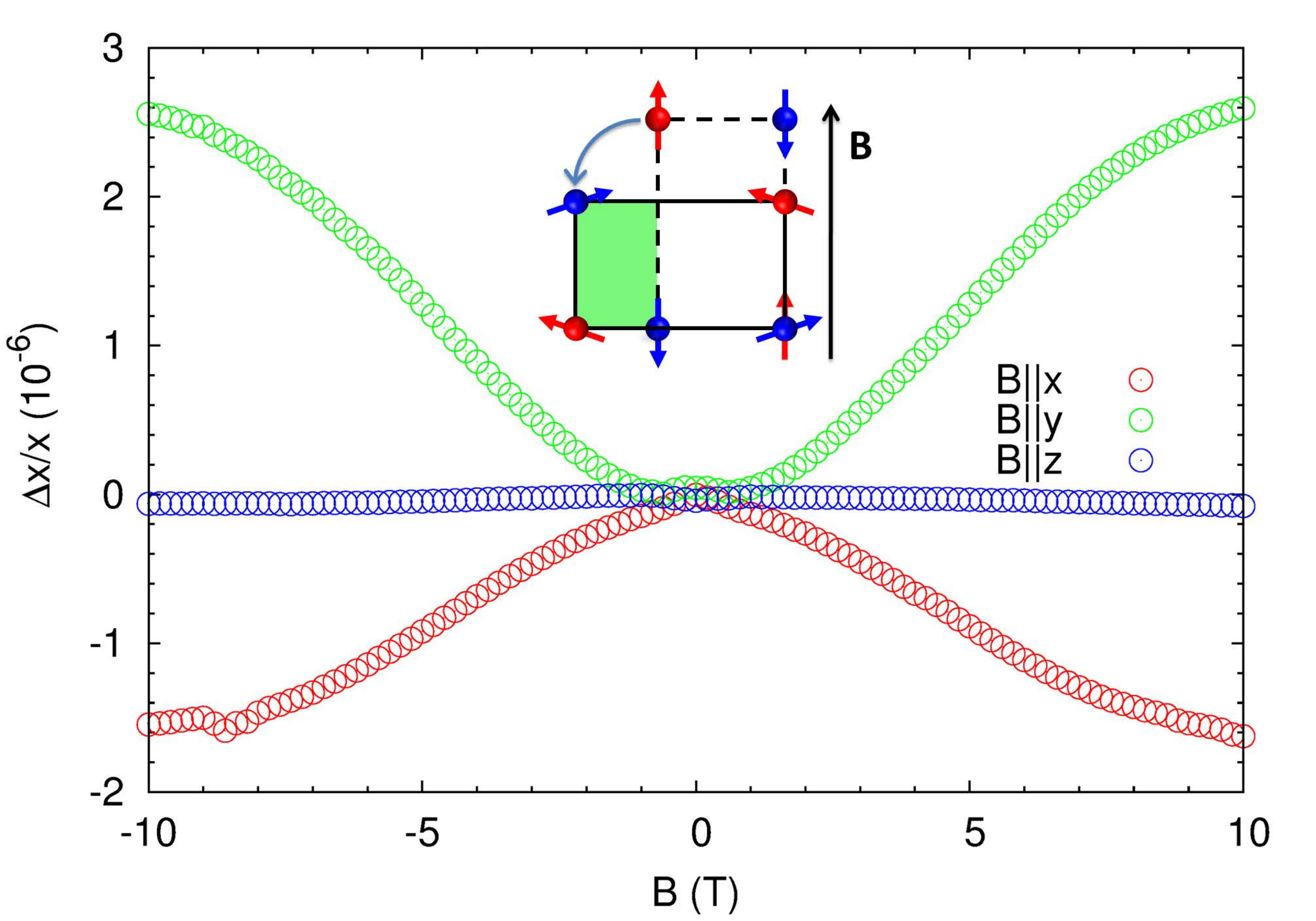}
    \caption{(color online) Forced magnetostriction at $T = 2$~K measured by dilatometry parallel to the Cu-O-Cu bond direction, $x$, in the CuO$_2$ plane. The field-induced change of the sample length along $x$ with magnetic field $\bf B$ applied parallel to the $x$, $y$, $z$ directions is plotted in red, green, and blue, respectively. Inset: Illustration of spin-orbit coupling induced magnetostriction for a single magneto-structural domain with ${\bm B} \parallel b$. Due to magnetostriction, the spin-flop transition induced by the field is associated with a realignment of the crystallographic unit cell (dashed line for ${\bm B} =0$, solid line for $\bm B \gtrsim 5$ T.) The orthorhombic distortion is exaggerated for clarity. \label{fig:3}}
\end{figure}

Since LD with radio-frequency coils is restricted to zero magnetic field, we used capacitative dilatometry as a complementary tool to investigate manifestations of forced magnetostriction in the antiferromagnetic state for $T = 2$~K. Figure~\ref{fig:3} shows the relative expansion of the $x$-axis along the Cu-O-Cu bonds, with magnetic field $\bm B$ along $x$, $y$ (in the CuO$_2$ planes), and $z$ (perpendicular to the planes).
For $\bm B \parallel y$ ($\bm B \parallel x$), $\Delta x/x$ is positive (negative), corresponding to expansion and contraction, respectively.
The resulting field-induced orthorhombic distortion of the crystal increases rapidly for small $\bm B$ and crosses over to a more gradual evolution for $\bm B_c \geq 5~\text{T}$ (defined as the inflection point in the $\Delta x/x$-versus-$\bm B$ relation).
The expansion for $\bm B \parallel z$ is close to zero. In stark contrast to classical antiferromagnets \cite{Felcher,Kalita}, there is no discernible field hysteresis of the forced magnetostriction which would indicate pinning of antiferromagnetic domain walls.

The dilatometry data indicate that the lattice expansion is coupled to the magnetic moment direction.
Related effects have been observed in other antiferromagnets including rare-earth magnets, where they can be understood as consequences of the spin-orbit interaction \cite{Doerr}.
Briefly, the spin-orbit interaction ties the spin direction to the orbital magnetization and hence to the shape of the valence electron cloud around the magnetic ions, which in turn is coupled to the lattice structure via crystalline electric fields. The small magnitude of the forced magnetostriction, compared to the manifestations of isotropic exchange striction discussed above, can then be attributed to the quenching of the spin-orbit interaction in the cuprates, where the magnetic dipole moment arises almost exclusively from the spin-1/2 of the Cu$^{2+}$ ions.
Nonetheless, the observed $g$-factor anisotropy of the Cu moments \cite{Walstedt} indicates a small residual orbital magnetization that can act as a source of magnetostriction.

The inset in Figure~\ref{fig:3} illustrates the spin-orbit mediated magnetostriction.
For $\bm B=0$, both neutron diffraction \cite{Burlet} and electron spin resonance \cite{Janossy3} find an equal population of domains with Cu spins oriented along the two orthogonal easy axes in the CuO$_2$ plane. Within each domain, the $a$ and $b$ axes are slightly different as a consequence of the spin-orbit interaction, but domain averaging results in a macroscopically tetragonal structure.
For increasing $\bm B \parallel y$, the Cu spins in the domain with spins pointing along $y$ flip by 90$^\circ$ to gain advantage of the Zeeman energy, whereas spins already along $x$ do not flip. The observed macroscopic expansion, $\Delta x/x$, is due to the slight orthorhombic distortion of each domain that is tied to the spin direction. For the same reason, $\Delta x/x$ is opposite in sign for $\bm B \parallel x$.
(The slight difference in the magnitudes of $\Delta x/x$ for $\bm B$ along $x$ and $y$ presumably arises from the uniaxial pressure along $x$ exerted by the sample holder, which increases the population of the domains with long axes $\perp x$.)
For $\bm B \geq \bm B_c$, most spins are oriented nearly perpendicular to the magnetic field, and the crystal structure is macroscopically orthorhombic. For larger fields, the gradual canting of the magnetic moments towards $\bm B$ is an additional source of magnetostriction, but this contribution is small because it is opposed by the large in-plane exchange interaction ($J \sim 100$ meV). The remarkable absence of field hysteresis may then be attributed to the approximate coincidence of magnetic and structural domain boundaries noted above. Since most structural mosaic blocks include a single magnetic domain, pinning of magnetic domain walls is largely suppressed.

We now discuss the relationship between the comprehensive picture of the magneto-structural coupling we have obtained to the transport properties of undoped and lightly doped cuprates reported earlier.
First, measurements of the magnon-mediated thermal conductivity of undoped, antiferromagnetic La$_2$CuO$_4$ have yielded low-temperature mean free paths in the range $\sim 100-150$~nm, \cite{Hess,Hofmann} somewhat lower than the magneto-structural domain size of $\sim 240$~nm inferred from our LD measurements on YBa$_2$Cu$_3$O$_{6.0}$, where thermal conductivity measurements have not yet been reported.
Since both experiments were carried out on different materials,
we regard the agreement as quite satisfactory. Our results suggest that magneto-structural domains limit the low-temperature heat conductivity mediated by magnons, and they provide a motivation for more detailed model calculations along these lines.

The spin-orbit mediated forced magnetostriction we identified in the antiferromagnetic state has the same ``$d$-wave'' symmetry ($i.e.$, positive parallel and negative perpendicular to the $\bm B$-field) and a similar crossover field ($\bm B_c \sim 5$~T) as the magnetoresistance in lightly doped antiferromagnetic YBa$_2$Cu$_3$O$_{6+\delta}$ \cite{Ando,Cimpoiasu}.
Our observations thus support models that ascribe the anomalous magnetoresistance to the magnetic field alignment of the orthorhombic magnetic domains. \cite{Cimpoiasu,Janossy1,Janossy2,Gomonaj,Janossy3}.
The orthorhombicity $a/b-1 = 4 \times 10^{-6}$ determined from the forced magnetostriction (Fig.~\ref{fig:3}) is somewhat smaller than the one estimated \cite{Cimpoiasu} on the basis of magnetoresistance data on YBa$_2$Cu$_3$O$_{6.25}$, but since this estimate is rather indirect, and both sets of measurements were taken on samples with different oxygen concentrations, the agreement is again quite satisfactory.
There is thus no need to invoke charge-stripe ordering in lightly doped YBa$_2$Cu$_3$O$_{6+\delta}$ to explain the magnetoresistance.
This is in accord with current knowledge of the phase diagram of this compound, where charge order only sets in at higher doping levels ($\delta \geq 0.5$) \cite{Keimer}.

In summary, the complementary combination of neutron Larmor diffraction and capacitative dilatometry has provided direct insight into the mesoscopic structure of the antiferromagnetic state in undoped YBa$_2$Cu$_3$O$_{6.0}$.
Our data allowed us to elucidate the magneto-structural coupling mechanisms and their influence on the heat and charge transport properties.
Based on the solid foundation we have laid here, our experimental approach can be straightforwardly applied to more highly doped cuprates, where domain structures associated with spin density wave, charge density wave, and ``nematic'' ordering phenomena and their influence on the macroscopic properties are subjects of intense current research and debate \cite{Keimer,Capati,Fernandes,Senthil,Zeljkovic,Kapitulnik,Shi,Achkar,LeTacon,Julien}.
More generally, we have established neutron Larmor diffraction as a versatile probe of antiferromagnetic and magneto-structural domain structures with sub-micrometer length scales, which opens up new perspectives for the investigation of a large variety of correlated-electron materials \cite{Dagotto}.

\begin{acknowledgments}
We are grateful to A. J\'anossy, S.P. Bayrakci, and J. Porras for discussions.
The work was supported by the DFG under Grant No. SFB/TRR 80.
B.N. acknowledges support by the \emph{Prospective Research Program} No. PBELP2-125427 of the Swiss NSF, and by the European Commission through the \emph{Research Infrastructures} action of the Capacities Programme, NMI3-II, Grant Agreement number 283883.
\end{acknowledgments}


\begin{thebibliography}{99}
\bibitem{note} $(\frac{1}{2}\:\frac{1}{2}\:5)$ is the only antiferromagnetic Bragg reflection of reasonable intensity that is accessible on TRISP. To determine the anisotropy of the magnetic correlation length, measurements on a larger crystal array are required.
\end{thebibliography}
\end{document}